# Element-Resolved Corrosion Analysis of Stainless-Type Glass-Forming Steels

**Authors:** M. J. Duarte[1,2,3,4±*], J. Klemm[1±], S. O. Klemm[1], K. J. J. Mayrhofer[1], M. Stratmann[1], S. Borodin[1], A. H. Romero[2,5ǂ], M. Madinehei[3], D. Crespo[3], J. Serrano[6], S. S. A. Gerstl[4§], P. P. Choi[4], D. Raabe[4], F. U. Renner[1¶*]

**Affiliations:**

[1] Department of Interface Chemistry and Surface Engineering, Max-Planck Institut für Eisenforschung GmbH, 40237 Düsseldorf, Germany.

[2] Departamento de Materiales, CINVESTAV-IPN, 76230 Queretaro, Mexico.

[3] Departament de Fisica Aplicada, Universitat Politècnica de Catalunya, 08860 Castelldefels, Spain.

[4] Department of Microstructure Physics and Alloy Design, Max-Planck Institut für Eisenforschung GmbH, 40237 Düsseldorf, Germany.

[5] Max-Planck-Institute für Mikrostrukturphysik, 06120 Halle, Germany

[6] ICREA- Universitat Politècnica de Catalunya, 08860 Castelldefels, Spain.

\* Correspondence to: j.duarte@mpie.de, f.renner@mpie.de

[§] Present address: ETH Zurich (EMEZ), 8093 Zurich, Switzerland. [ǂ] Present address: Physics Department, West Virginia University, Morgantown, WV, 26506-6315, USA. [¶] Present address: Instituut Voor Materiaalonderzoek, Universiteit Hasselt, 3590 Diepenbeek, Belgium.

[±] These authors contributed equally to this manuscript.

**Abstract**:

Ultrathin passive films effectively prevent the chemical attack of stainless steel grades in corrosive environments; their stability critically depends on the interplay between structure and chemistry of the constituents Fe-Cr-Mo. In particular, nanoscale inhomogeneities along the surface can have a tremendous impact on material failure, but are yet barely understood. Addressing a stainless-type glass-forming $Fe_{50}Cr_{15}Mo_{14}C_{15}B_6$ alloy and utilizing a combination of complementary high-resolution analytical techniques, we relate near-atomistic insight into different gradual nanostructures with time- and element-resolved dissolution behavior. The progressive elemental segregation on the nanoscale is followed in its influence on the concomitant degree of passivity. A detrimental transition from Cr-controlled passivity to Mo-controlled breakdown is dissected atom-by-atom demonstrating the importance of nanoscale knowledge for understanding corrosion.

**Main Text:**

The environmental degradation of metals by chemical reactions in humid atmospheres or electrolytes produces huge worldwide annual losses. In applications where corrosion is a critical factor, typically resistant poly-crystalline Ni-based materials or relatively cheaper Fe-based steels are employed. In these stainless materials Cr plays a vital role in establishing protective passive films of a few atomic layers *(25-28)*, which show improved stability by combination with Mo *(29,30)*. Glassy alloys, including amorphous steels, have gained attention, among other due to their extraordinary corrosion and pitting resistance at comparatively low Cr contents *(6-9)*. Additionally, an improved mastering of production *(10)* opens promising applications ranging from structural gadgets to biocompatible implant materials *(11-13)*. Moreover, the chemical and structural homogeneity of amorphous steel alloys have been recognized as an important asset for fundamentally addressing the influence of alloying elements on the macroscopic behavior such as corrosion *(7,14,15)*. The metastable glassy structure can be devitrified by gentle annealing *(16)*, which does enable well-defined intermediate states of nanoscale elemental inhomogeneity. Although inhomogeneous on the nanoscale, the annealed structures are homogeneous on the macro-scale. Microscopic insights of bulk material can thus be related to macroscopic performance.

Addressing the nanoscopic passive films of complex stainless steels remains a challenge. Large progress in understanding the laterally-averaged structure and chemistry of the ultrathin passive films has been achieved by surface analysis techniques *(6,17,18)*. Despite their importance, however, the influence of nanoscale inhomogeneities on the film reactions such as crystallographic defects, small-scale inclusions, precipitates, grain boundaries, or purely chemical variations along surfaces, remains largely unresolved and it is key for advancing modern materials development and corrosion research *(19-21)*. The passive films effectively change or block the dissolution of metallic materials in their specific environment. A sensitive multi-elemental online detection of dissolved species during their build-up or steady-state regime, together with a finely resolved knowledge of the underlying substrate, makes this effect visible and allows insight into the conditions it is based on.

We provide a high-resolution nanoscale and atomistic view on the relation between bulk and surface microstructure and time-resolved corrosion performance of the stainless-type amorphous and nanocrystalline alloy $Fe_{50}Cr_{15}Mo_{14}C_{15}B_6$ *(22)*. Amorphous ribbons were prepared by melt spinning and crystallized by thermal annealing at the selected temperatures for 20 min. The glass transition ($T_g$ = 550 °C), the onset of the first and second crystallization ($T_{x1}$ = 602 °C, $T_{x2}$ = 641 °C), and the melting temperatures ($T_m$ = 1112 °C) of the amorphous alloy were obtained by Differential Scanning Calorimetry and serve as an indicator of the structural transformations within the alloy.

Atom Probe Tomography (APT) *(23)* reconstructions are presented in Fig. 1A (and Figs. S1,2). Fe, Cr and Mo, are presented in green, blue, and red colors respectively. Boron and carbon were omitted for clarity. The lower images correspond to 2-5 nm slices from the atom maps. Isoconcentration surfaces of Cr and Mo are plotted to highlight regions of highest chemical gradients. The average compositions of the different regions are listed in Table 1. The as-quenched ribbons show the broad diffuse X-ray peaks (Fig. 1B) characteristic of an amorphous structure, while a uniform elemental distribution by APT confirms their chemical homogeneity. At 620 °C Cr-rich regions of 12 nm average size are formed by elemental redistribution (Fig 1A). Correlated analyses with X-ray Diffraction (XRD) and Transmission Electron Microscopy (TEM) indicate the presence of the carbide phase $M_{23}(C,B)_6$ (with M=Fe,Cr,Mo). Mo atoms segregate from these carbides to the amorphous matrix forming additional regions enriched in Mo (red areas at 620 °C) until finally a crystalline η-Fe$_3$Mo$_3$C phase is observed at 650 °C. M$_3$B and a solid solution Fe-Cr-Mo are present as intermediate phases only in the 620-650 °C temperature range. The fully crystallized sample (at 800 °C) shows Mo-rich and Cr-rich phases of up to 50 nm in size (see also Fig. S3). The chemical composition in the respective regions becomes thermodynamically stable with increasingly sharp boundaries. Most important, a percolation of both crystalline phases throughout the fully crystallized sample is observed.

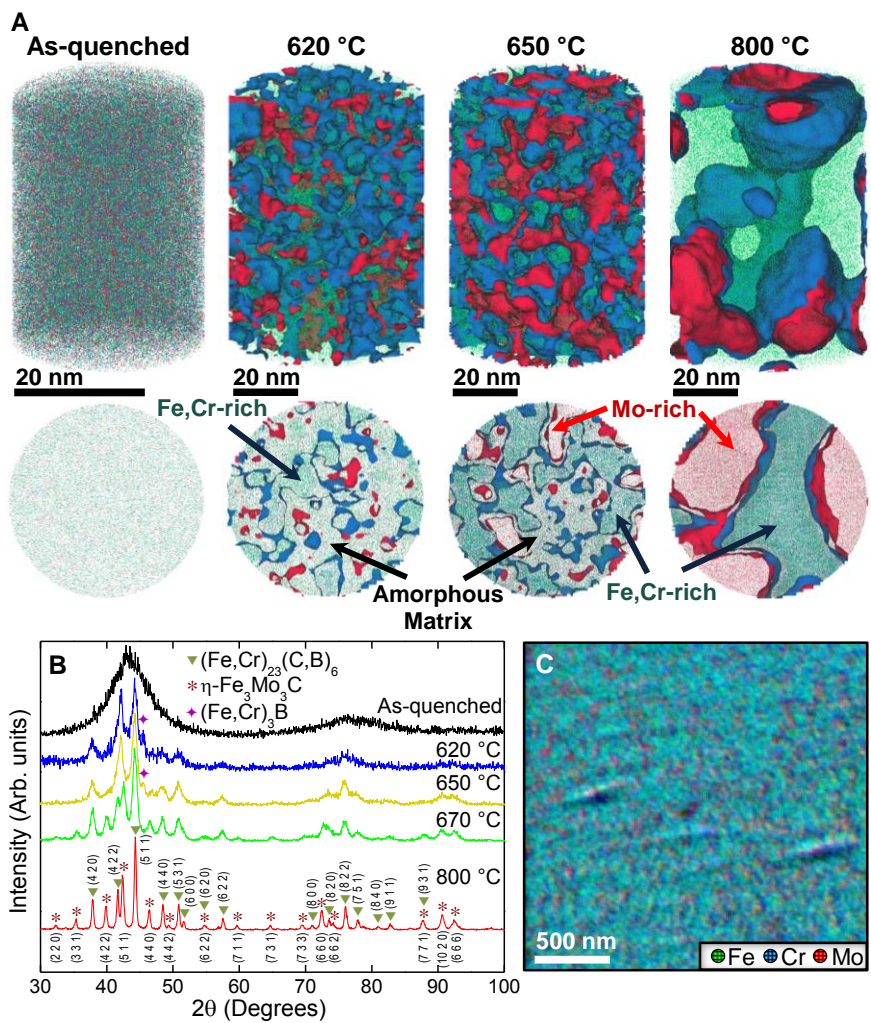

**Fig. 1. Microstructure and surface analysis.** (**A**) APT reconstructions showing the distribution of the metallic elements (Fe: green •, Cr: blue •, Mo: red •) in the as-quenched, partially and fully crystallized alloys. Isoconcentration surfaces of 22 at.% Mo in red and 16 at.% Cr in blue highlight three different regions in the samples annealed for 20 min at 620 °C and 650 °C. 32 at.% Mo and 15 at.% Cr isoconcentration surfaces are plotted for the crystallized alloy (60 min at 800 °C). Lower images are 2-5 nm thick virtual slices of the respective reconstructions. (**B**) XRD plots indicating the formation of the different phases. (**C**) Overlap of the individual Fe, Cr and Mo SAEM maps after crystallization at 800 °C.

Bombardment with Ar ions essentially cleaned the surfaces from contaminations or any reaction layers formed during annealing (Fig. S4). The elemental distribution of the fresh surface of the fully crystalline sample in the Scanning Auger Electron Microscopy (SAEM) map (overlaid map in Fig. 1C) closely corresponds to the distribution displayed in the APT slice in Fig.1A. Although some sputter ripples emerge in the surface morphology, sputtering is thus able to produce an almost ideal cut of the bulk material and the detailed bulk microstructure information is maintained at the surface. Freshly sputtered starting surfaces were used to ensure reproducibility of the electrochemical measurements and surface analysis.

**Table 1. Representative concentration** of the metallic components (in at. %) in the different regions determined by APT and SFC-ICP-MS.

| | | | APT | | | | | | | | |
|---|---|---|---|---|---|---|---|---|---|---|---|
| Alloy | Am | 550°C | 620 °C | | 650 °C | | | 670 °C | | 800 °C | |
| Zone [in Fig. 1A] | | | Cr-rich | Matrix | Cr-rich | Mo-rich | Matrix | Cr-rich | Mo-rich | Cr-rich | Mo-rich |
| % Fe | 59 | 59 | 61 | 58 | 65 | 44 | 59 | 65 | 44 | 65 | 42 |
| % Cr | 19 | 18 | 26 | 19 | 23 | 9 | 19 | 23 | 8 | 22 | 7 |
| % Mo | 22 | 23 | 13 | 23 | 12 | 47 | 22 | 12 | 48 | 13 | 51 |

| | | | ICP – MS | | | | | | | | |
|---|---|---|---|---|---|---|---|---|---|---|---|
| Alloy | Amorphous (Am) | | | 550 °C | | | 620 °C | | | 670°C | 800°C |
| Region [in Fig.2] | a | b | c | a | b | c | a | b | c | a | a |
| % Fe | 65 | 68 | 63 | 68 | 72 | 63 | 71 | 70 | 64 | 56 | 53 |
| % Cr | 13 | 16 | 20 | 11 | 14 | 19 | 7 | 10 | 17 | 7 | 4 |
| % Mo | 22 | 16 | 17 | 21 | 14 | 18 | 22 | 20 | 19 | 37 | 43 |

Micro-electrochemical corrosion tests enabled a large number of tests on one same sample and showed an excellent reproducibility (24). Depending on the thermal history, the electrochemical corrosion behavior reveals a transition from an extended passive state limited by transpassive Cr dissolution to an early breakdown governed by transpassive Mo dissolution (Fig. 2A in 0.1M $H_2SO_4$ and S5 in 0.1 M HCl). The curves obtained are located between the reference behavior of pure Mo and pure Cr. In region (a) of Fig. 2A oxidizing positive currents indicate the formation of metal ions. For the alloys crystallized above 670 °C the steep increase in current is linked to strong metal dissolution. Below 650 °C the annealed and the pristine amorphous alloys show a transition into a further low-current passive region (b). In region (c) the passive film breaks down close to the transpassive Cr dissolution of the pure Cr foil. Interestingly, the partially nanocrystalline samples below 620 °C show only a moderately larger passive current density. The transition peak (a) is more pronounced and the final breakdown occurs at the same potential compared to the amorphous alloy. The transition in breakdown behavior appears between 620 °C and 670 °C. The fully nanocrystalline sample (800 °C) closely resembles the behavior of pure Mo, even though APT analysis indicates 6-7 at.% of Cr in the respective Mo-rich phase.

Of particular interest for corrosion performance are the individual elemental contribution and the effectiveness of the passivating element Cr and of Mo (7,25,26). Element-resolved dissolution data was measured online using a Scanning Flow Cell (SFC) with an Inductively Coupled Plasma Mass Spectrometer (ICP-MS) (27) and illustrated for three selected states in Fig. 2B-G. The measured concentrations were converted into current density by application of Faraday's law (27-29). The sum of all three elemental dissolution current densities, $i_{Diss,Me}$ with Me = Fe, Cr, Mo, (grey dashed lines in Fig. 2B-D) shows a good agreement with the measured current density $i_m$ for all three samples. The deviation reflects the growth of the passive film of about 1-2 nm in thickness and dissolution of carbon and boron. The initial peak indicates the transition of a native oxide film, which is unstable in the electrolyte, towards this passive state. Compared with the bulk composition the element-specific dissolution stoichiometry (molar fractions) reveals significant changes (Fig. 2E-G). For the amorphous sample the main species dissolved in the lower range of potentials up to 700 mV is Fe with only sub-stoichiometric dissolution of Cr. The time-resolved enrichment of Cr on the surface can thus directly be followed. At higher potentials in the passive range, both Mo and Cr dissolve in equivalent ratio reflecting the substrate composition. In the sample annealed at 620 °C a higher contribution of Fe and Mo is observed. Cr dissolves then stoichiometrically only at the breakdown. The formation of Cr-rich crystals, and consequently Cr-depleted regions, leads to the increase in the current density. The passive layer formation is in both presented stages accompanied by a large etching effect which is enhanced by the detected co-dissolution of Cr. A relatively thick layer of the amorphous and the partially crystalline surface is removed (about 20 nm and 30 nm in average, respectively) as can be deduced from the elemental dissolution data. A thicker and significantly Cr-enriched adjusted passive film is thus formed after immersion in the electrolyte. Independent X-ray Photoelectron Spectroscopy (XPS) measurements (Fig. S6) confirm the increase in film thickness to about 4 nm and also reveal that the thickness and composition of the respective passive films resemble closely to the reported films on binary Fe-Cr alloys or stainless steels. In the fully crystalline alloy (800°C) an increasingly strong contribution of Mo is detected, with Cr being distinctly minor. The dissolution stoichiometry is thus quickly dominated by the Mo-rich phase and the chemical stability of the system is compromised with the formation of the percolating Cr-depleted structure. The individual amounts of metals dissolved in the specific potential regions are given in Table 1. The ratio of Fe and Mo measured by APT in the Mo-rich phase agrees very well to the dissolution of the nanocrystalline material during the potential sweeps. This supports that the electrochemical behavior in the crystallized alloys is indeed dominated by the Mo-rich and severely Cr-depleted phase, which is corroding already at low potentials. In parallel to the growth of the adapted passive film the associated selective etching develops a severe surface roughening on the nanometer length-scale of the formed chemical inhomogeneities and crystallites. Fig. 3A shows an Atomic Force Microscopy (AFM) image of the surface morphology after initial corrosion of the partially crystalline state (annealed at 620 °C). The final trans-passive breakdown of the layer is here not compromised by the developing defects in the surface morphology and the partial crystallization. The completely amorphous sample shows no difference in roughness before and after formation of the passive layer confirming its homogeneous nature. Only at high temperatures a more pronounced chemical partitioning causes the mechanism and the breakdown potential to change.

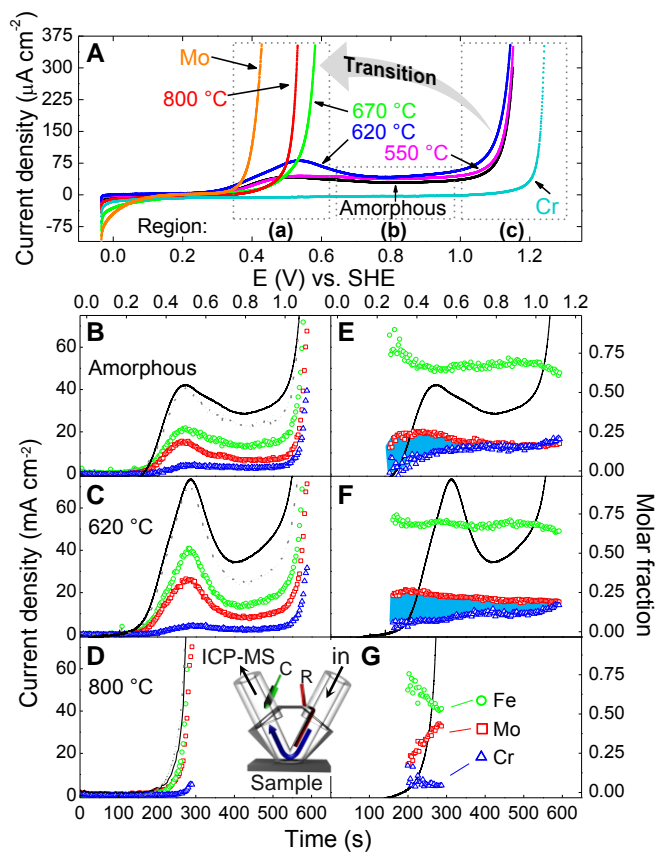

**Fig. 2. Electrochemical behavior and element resolved online analysis** in 0.1M $H_2SO_4$. (**A**) Potentiodynamic polarization curves (2 mV/s in the positive direction) of amorphous, partially (620°C) - and fully nanocrystalline (800°C) $Fe_{50}Cr_{15}Mo_{14}C_{15}B_6$ together with pure Cr and Mo. (**B-D**) Online-ICP-MS dissolution profiles converted into dissolution current densities ($i_{Diss,Me}$; Me = Fe, Cr, Mo with green, blue and red symbols respectively, and $i_{Diss,total}$ = sum of individual $i_{Diss,Me}$, grey dashed line) for the (**B**) amorphous, (**C**) partially nanocrystalline (**D**) and fully nanocrystalline alloy. Potential sweep data, $i_m$, as black line. (**E-F**) Dissolution stoichiometry expressed as molar fractions of Fe, Cr and Mo, for the (**E**) amorphous, (**F**) partially nanocrystalline and, (**G**) fully nanocrystalline alloys. The difference between Mo and Cr dissolution is highlighted in blue. The inset shows a schematic figure of the SFC-ICP-MS.

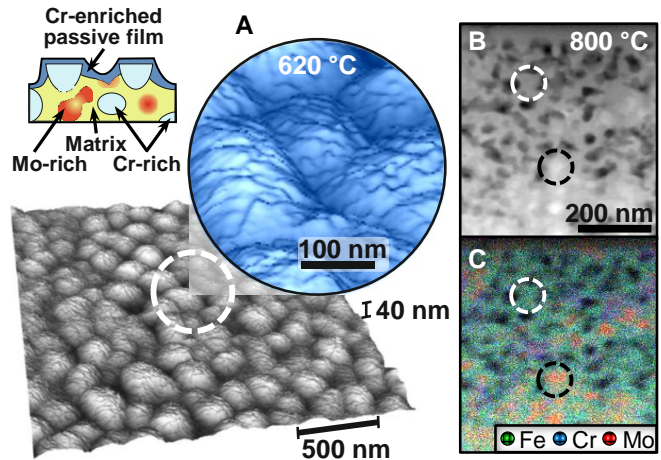

**Fig. 3**. **Surface analysis and cross section after galvanostatic corrosion.** (**A**) AFM images and sketch of a rough surface after corrosion of a sample annealed at 620°C for 20 min. (**B**) HAADF-STEM micrograph of the fully crystallized alloy after polarization for 4000s and (**C**) corresponding EDX map showing the Mo, Fe and Cr distributions in the porous layer and the bulk. Mo-rich and Cr-rich areas are bordered by black and white dashed circles, respectively.

In the case of percolating microstructures, e.g. the alloy crystallized at 800 °C, continuous selective dissolution can be sustained as shown in Fig. 3B and C. Chemical segregation is independently confirmed as variations in contrast in the high angle annular dark field (HAADF) Scanning Transmission Electron Microscopy (STEM) image of Fig. 3B. Quantitative results from the darker regions obtained by local Energy Dispersive X-Ray (EDX) analysis show an enrichment in Cr, up to 22 at.%, in the porous layer (white dashed circle). Mo-rich areas, such as the one surrounded in black, contain around 39 at. % Mo, with a Cr concentration below 10 at. %. As can be seen more clearly in the Mo-Fe-Cr overlay EDX map (Fig. 3B), only a small concentration of Mo is present in the porous layer, confirming the dissolution of a percolating Mo-rich phase.

The complexity of most applied materials often precludes a thorough understanding of macroscopic behavior, including corrosion. Here, we followed the role and fate of the individual elements in corrosion of a complex multi-element material. First, the 3-dimensional chemical distribution revealed by APT discloses the increasing Cr partitioning of the alloy at different temperatures which is essential for understanding the breakdown of passivity. The Cr-depletion is first occurring homogeneously throughout the remaining matrix, i.e. no direct sensitization is observed. The Cr-depleted η-$Fe_3Mo_3C$ phase does not form surrounding, even not always attached to, the Cr-enriched crystallites. Approaching the fully crystallized state at 700-800 °C the η-phase is percolated throughout the alloy.

A second major point is the online ICP-MS dissolution analysis following the resulting time-resolved enrichment of Cr on the surface, which drives passivation. The initial formation peak before the Cr passive range shows substoichiometric dissolution of Cr together with increased Mo dissolution. The partially crystalline samples exhibit higher initial dissolution peaks before passivity is established. The subsequent low-current plateau at higher Cr dissolution compared to the initial passivation peak, and slightly higher passive current compared with the amorphous, is established and marking a change in the mechanism during the potential ramp. As Cr is also co-dissolved, a more significant surface etching (1-3 nm) takes place with the passive film formation, which may help to overgrow a structurally inhomogeneous substrate. After breakdown the (previously) passive microstructures dissolve with bulk stoichiometry. This transpassive breakdown process occurs at lower potentials for the analyzed alloys compared to pure Cr, thus pinpointing a further Cr co-dissolution effect. The early breakdown of the fully nanocrystalline state is associated with a very high Mo and Fe dissolution. This drastic change in elemental dissolution stoichiometry at a high degree of crystallization (700-800 °C) reflects a sudden passage from a passive state with a Cr-controlled transpassive breakdown to a Mo-controlled breakdown. The transition which represents a severe detrimental effect on the corrosion properties is due to the pronounced chemical segregation with formation of a percolating Mo-rich nanostructure which is preferentially attacked and dominating the electrochemical corrosion behavior.

As a general remark, the structure of passive films on stainless steels has been recognized as an amorphous-crystalline composite *(30)*. On the other hand, the exact nanometer-scale lateral chemical composition of any ultrathin passive film has not been resolved so far. Amorphous steels show a low threshold *(31)* to develop a protective film of 4-5% of Cr compared to 12-13% of Cr in crystalline stainless steel alloys *(32)*. The origin of this lower threshold remains disputed and was variously attributed to either the amorphous substrate facilitating an entirely amorphous passive film structure, or to a more homogeneous Cr distribution. With the limited knowledge on the passive film itself element-resolved solution analysis becomes very important.

Finally, beyond the presented peculiar amorphous and nanocrystalline alloy, interesting parallels to commercial polycrystalline stainless steels can be drawn. The respective passive films of both are of the same nature i.e. they have the same thickness and average composition. High-performance austenitic stainless steel grades such as AISI type 316L, 904L, or 654 SMO® contain a relativly high amount of 2-7 wt. % of Mo and are employed where corrosion behavior is critical. Passive films are not static but adapt to changes in the environment and are partially able to heal out defective regions. Local variations in Cr for example such as around weld precipitations, tiny impurity phases, or sensitized grain boundaries are discussed as a major cause of film breakdown in corrosion of stainless steels, while the role of very small nanoscale fluctuations has been barely addressed. Amorphous and nanocrystalline steels with their excellent homogeneity on the macroscale combined with the possibility to obtain well-defined gradual (structural and chemical) inhomogeneities on the nanoscale are possible model materials for understanding real world steels.

**Acknowlegments:** We acknowledge Dr. Aleksander Kostka providing TEM, STEM and EDX measurements and helpful discussions. For assistance in SAEM we thank Dr. Patrick Keil. Andrea Mingers is acknowledged for ICP-MS measurements. MJD, JS and DC acknowledge financial support from Micinn MAT2010-14907, Generalitat de Catalunya 2009SGR01225 and 2009SGR01251. MJD and AHR acknowledge support from Conacyt, Mexico, through projects PPROALMEX-DAAD-Conacyt and 152153. AHR was supported by the Marie-Curie Intra-European Fellowship. FUR acknowledges support by DAAD PPP 0811941.